\documentclass[12pt]{iopart}

\usepackage{epsfig}

\newcommand{\be}{\begin{equation}}
\newcommand{\ee}{\end{equation}}
\newcommand{\bea}{\begin{eqnarray}}
\newcommand{\eea}{\end{eqnarray}}

\begin{document}

\title{Charmonium dynamics in heavy ion collisions}

\author{O.~Linnyk$^1$, E.~L.~Bratkovskaya$^1$, W.~Cassing$^2$, H.~St\"ocker$^{1,3}$}

\address{$^1$
Frankfurt Institute for Advanced Studies, %
 Ruth-Moufang-Str. 1, %
 60438 Frankfurt am Main, %
 Germany}%
\address{$^2$
Institut f\"ur Theoretische Physik, %
  Universit\"at Giessen, %
  Heinrich--Buff--Ring 16, %
  35392 Giessen, %
  Germany}%
\address{$^3$
Institut f\"ur Theoretische Physik, %
 Johann Wolfgang Goethe University, %
 Max-von-Laue-Str. 1, %
 60438 Frankfurt am Main, %
 Germany%
}

\ead{linnyk@fias.uni-frankfurt.de}


\begin{abstract}
Applying the HSD transport approach to charmonium dynamics within
the `hadronic comover model' and the `QGP melting scenario', we
show that the suppression pattern seen at RHIC cannot be explained
by the interaction with baryons, comoving mesons and/or by color
screening mechanism. The interaction with hadrons in the late
stages of the collision (when the energy density falls below the
critical) gives a sizable contribution to the suppression. On the
other hand, it does not account for the observed additional
charmonium dissociation and its dependence on rapidity. Together
with the failure of the hadron-string models to reproduce high
$v_2$ of open charm mesons, this suggests strong pre-hadronic
interaction of $c \bar c$ with the medium at high energy
densities.
\end{abstract}




\maketitle


\section{$J/\Psi$ production vs suppression in different theoretical scenarios}

The microscopic Hadron-String-Dynamics (HSD) transport
calculations (employed here) provide the correct space-time
geometry of the nucleus-nucleus reaction and a rather reliable
estimate for the local energy densities achieved.  The energy
density $\varepsilon({\bf r};t)$ -- which is identified with the
matrix element $T^{00}({\bf r};t)$ of the energy momentum tensor
in the local rest frame at space-time $({\bf r},t)$ -- reaches as
high as 30~GeV/fm$^3$ in a central Au+Au collision at $\sqrt{s}$ =
200 GeV~\cite{Olena2}.

According to present knowledge, the charmonium production in
heavy-ion collisions, {\it i.e.}  $c\bar{c}$ pairs,  occurs
exclusively at the initial stage of the reaction in primary
nucleon-nucleon collisions. The parametrizations of the total
charmonium cross sections ($i = \chi_c, J/\Psi, \Psi^\prime$) from
$NN$ collisions as a function of the invariant energy $\sqrt{s}$
used in this work are taken
from~\cite{Cass99,Cass00,brat03,Cass01}. We recall that (as in
Refs. \cite{brat03,Cass01,Geiss99,Cass97,CassKo}) the charm
degrees of freedom in the HSD approach are treated perturbatively
and that initial hard processes (such as $c\bar{c}$ or Drell-Yan
production from $NN$ collisions) are `precalculated' to achieve a
scaling of the inclusive cross section with the number of
projectile and target nucleons as $A_P \times A_T$ when
integrating over impact parameter. For fixed impact parameter $b$
the $c\bar{c}$ yield then scales with the number of binary hard
collisions $N_{bin}$ ({\it cf.} Fig. 8 in Ref.~\cite{Cass01}).

In the QGP `threshold scenario', e.g the geometrical Glauber model
of Blaizot et al.~\cite{Blaizot} as well as the percolation model
of Satz~\cite{Satzrev}, the QGP suppression `(i)' sets in rather
abruptly as soon as the energy density exceeds a threshold value
$\varepsilon_c$, which is a free parameter. This version of the
standard approach  is motivated by the idea that the charmonium
dissociation rate is drastically larger in a quark-gluon-plasma
(QGP)  than in a hadronic medium~\cite{Satzrev}.

On the other hand, the extra suppression of charmonia in the high
density phase of nucleus-nucleus collisions at SPS
energies~\cite{NA50aa,NA50b,NA50a,NA60} has been attributed to
inelastic comover scattering ({\it
cf.}~\cite{Cass99,Cass00,Cass97,Olena,Capella,Vogt99,Gersch,Kahana,Spieles,Gerland}
and Refs. therein) assuming that the corresponding $J/\Psi$-hadron
cross sections are in the order of a few
mb~\cite{Haglin,Konew,Ko,Sascha}. In these models `comovers' are
viewed not as asymptotic hadronic states in vacuum but rather as
hadronic correlators (essentially of vector meson type) that might
well survive at energy densities above 1 GeV/fm$^3$. Additionally,
alternative absorption mechanisms  might play a role, such as
gluon scattering on color dipole states as suggested in
Refs.~\cite{Kojpsi,Rappnew,Blaschke1,Blaschke2} or charmonium
dissociation in the strong color fields of overlapping
strings~\cite{Geiss99}.

The explicit treatment of initial $c\bar{c}$ production by primary
nucleon-nucleon collisions and the implementation of the comover
model - involving a single matrix element $M_0$ fixed by the data
at SPS energies - as well as the QGP threshold scenario in HSD
were explained in Ref.~\cite{Olena2,Olena} (see Fig.~1 of
Ref.~\cite{Olena} for the relevant cross sections). We recall that
the `threshold scenario' for charmonium dissociation is
implemented as follows: whenever the local energy density
$\varepsilon(x)$ is above a threshold value $\varepsilon_j$ (where
the index $j$ stands for $J/\Psi, \chi_c, \Psi^\prime$), the
charmonium is fully dissociated to $c + \bar{c}$. The default
threshold energy densities adopted are $\varepsilon_1 = 16$
GeV/fm$^3$ for $J/\Psi$, $\varepsilon_2 = 2$ GeV/fm$^3$ for
$\chi_c$, and $\varepsilon_3 =2 $ GeV/fm$^3$ for $ \Psi^\prime$.
Two more scenarios were implemented similarly to the `comover
suppression' and the `threshold melting' by adding the only
additional assumption -- that the comoving mesons (including the
$D$-mesons) exist only at energy densities below some cut-energy
density $\epsilon _{cut}$, which is a free parameter. We set
$\epsilon _{cut}=1$~GeV/fm$^3$, {\it i.e.} of the order of
critical energy density.


\section{Comparison to data}

\begin{figure}
\centerline{ \psfig{figure=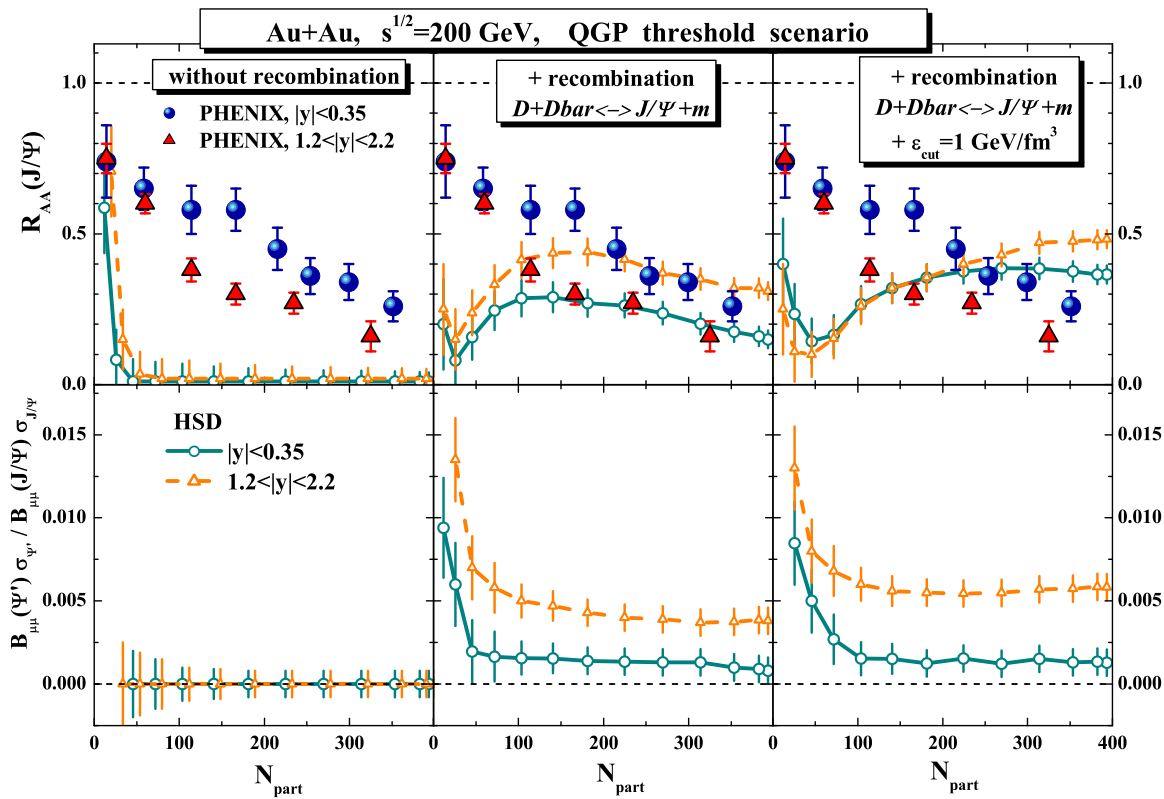,width=\columnwidth}}
\caption{The $J/\Psi$ nuclear modification factor $R_{AA}$ for
$Au+Au$ collisions at $\sqrt{s}=200$~GeV as a function of the
number of participants $N_{part}$ in comparison to the data from
[10] for midrapidity (full circles) and forward rapidity (full
triangles). HSD results for the QGP `threshold melting' scenarios
are displayed in terms of the lower (green solid) lines for
midrapidity $J/\Psi$'s ($|y| \le 0.35$) and in terms of the upper
(orange dashed) lines for forward rapidity ($1.2 \le y \le 2.2$)
within different recombination scenarios (see text). The error
bars on the theoretical results indicate the statistical
uncertainty due to the finite number of events in the HSD
calculations. Predictions for the ratio $B_{\mu \mu} (\Psi ')
\sigma _{\Psi'} / B_{\mu \mu} (J/\Psi) \sigma _{J/\Psi} $ as a
function of the number of participants $N_{part}$ are shown in the
lower set of plots. The figure is taken from~\cite{Olena2}.}
\label{RHICthreshold}
\end{figure}

\begin{figure}
\centerline{\psfig{figure=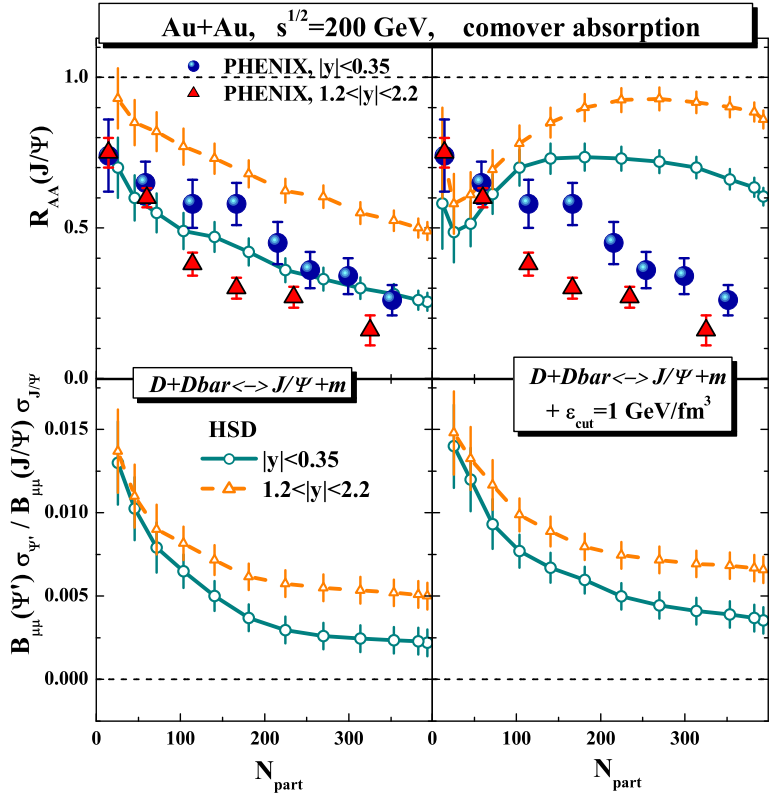,width=0.65\columnwidth}}
\caption{Same as Fig.~4 for the `comover absorption scenario'
including the charmonium reformation channels without cut in the
energy density (l.h.s.) and with a cut in the energy density
$\epsilon _{cut} = 1$~GeV/fm$^3$ (see text for details). The
figure is taken from~\cite{Olena2}. } \label{RHICcomover}
\end{figure}

In the following, we compare our calculations to the experimental
data at the top RHIC energy of $\sqrt{s}=200$~GeV. We recall that
the experimentally measured nuclear modification factor $R_{AA}$
is given by
\begin{equation}
R_{AA}=\frac{ d N (J/\Psi) _{AA} / d y  }{ N_{coll} \cdot d N
(J/\Psi) _{pp} / d y },
\end{equation}
where $d N (J/\Psi) _{AA} / d y $ denotes the final yield of
$J/\Psi$ in $A A$ collisions, $d N (J/\Psi) _{pp} / d y$ is the
yield in elementary $p p$ reactions, $N_{coll}$ is the number of
binary collisions.

Due to very high initial energy density reached (corresponding to
$T \gg 2 T_c$), in the threshold melting scenario all initially
created $J/\Psi$, $\Psi'$ and $\chi _c$ mesons melt. However, the
PHENIX collaboration has found that at least 20\%  of $J/\Psi$ do
survive at RHIC~\cite{PHENIX}. Thus, the importance of charmonium
recreation is shown again. In HSD, we account for $J/\Psi$
recreation via the $D  \bar D$ annihilation processes as explained
in detail in~\cite{Olena2,Olena}. Note that in our approach, the
cross sections of charmonium recreation in $D + \bar D \to J/\Psi
+ meson$ processes is fixed by detailed balance from the comover
absorption cross section $J/\Psi + meson \to D + \bar D$. But even
after both these processes are added to the threshold melting
mechanism, the centrality dependence of the $R_{AA} (J/\Psi)$
cannot be reproduced in the `threshold melting' scenario,
especially in the peripheral collisions (see
Fig.~\ref{RHICthreshold}). This holds for both possibilities: with
(r.h.s. of Fig.~\ref{RHICthreshold}) and without (center of
Fig.~\ref{RHICthreshold}) the energy density cut $\epsilon_{cut}$,
below which $D$-mesons and comovers exist and can participate in
$D + \bar D \leftrightarrow J/\Psi + meson$ reactions.

Comover absorption scenarios give generally a correct dependence
of the yield on the centrality. If an existence of D-mesons at
energy densities above 1 GeV/fm$^3$ is assumed, the amplitude of
suppression of $J/\Psi$ at mid-rapidity is also well reproduced
(see the line for `comover without $\epsilon_{cut}$' scenario in
Fig.\ref{RHICcomover}, l.h.s.). Note that this line correspond to
the prediction made in the HSD approach in~\cite{brat04}. 
On the other hand, the rapidity dependence of the comover result
is wrong, both with and without $\epsilon _{cut}$. If hadronic
correlators exist only at $\epsilon < \epsilon _{cut}$, comover
absorption is insufficient to reproduce the $J/\Psi$ suppression
even at mid-rapidity (see Fig.~\ref{RHICcomover}, r.h.s.). But its
contribution to the charmonium suppression is, nevertheless,
substantial. The difference between the theoretical curves marked
`comover + $\epsilon _{cut}$' and the data shows the maximum
possible supression that can be attributed to a deconfined medium.

%
%
%

\section{Summary}

The formation and suppression dynamics of $J/\Psi$, $\chi_c$ and
$\Psi^\prime$ mesons has been studied within the HSD transport
approach for $Au+Au$ reactions at $\sqrt{s}$ = 200 GeV. Two
currently discussed models, i.e. the 'hadronic comover absorption
and reformation' model as well as the 'QGP threshold melting
scenario' have been compared to the available experimental data.

We find that both `comover absorption' and `threshold melting'
fail severely at RHIC energies~\cite{Olena2}. The failure of the
'hadronic comover absorption' model goes in line with its
underestimation of the collective flow $v_2$ of leptons from open
charm decay as investigated in Ref.~\cite{brat05}. This suggests
that the dynamics of $c, \bar{c}$ quarks at this energy are
dominated by strong pre-hadronic interaction of charmonia with the
medium in strong QGP (sQGP), which  cannot be modeled by
`hadronic' scattering or described appropriately by color
screening.

On the other hand, the interaction of $J/\Psi$ with hadrons in the
late stages of the collision (when the energy density falls below
the critical) gives a sizable contribution to its suppression.
This contribution should not be neglected, when comparing possible
models for QGP-induced charmonium suppression to the experimental
data.

\vspace{0.5cm}

\bibliographystyle{h-physrev3}
\bibliography{HSDcharm}

\end{document}